\documentclass[aps,prb,longbibliography,reprint,showpacs,superscriptaddress]{revtex4-1}

\usepackage{graphicx}
\usepackage{color}
\usepackage{txfonts}
\usepackage[colorlinks, linkcolor=blue, anchorcolor=blue, citecolor=blue, urlcolor=blue]{hyperref}

\newcommand{\slro}{Sr$_{1.5}$La$_{0.5}$RhO$_{4}$}
\newcommand{\slrox}{Sr$_{2-x}$La$_{x}$RhO$_{4}$}
\newcommand{\sro}{Sr$_{2}$RhO$_{4}$}
\newcommand{\lsro}{LaSrRhO$_{4}$}

\begin{document}


\title{Single crystal growth and physical properties of 50\% electron doped rhodate Sr$_{1.5}$La$_{0.5}$RhO$_{4}$}



\author{Z. W. Li}
\affiliation{Max-Planck-Institute for Chemical Physics of Solids, N\"{o}thnitzer Str. 40, D-01187 Dresden, Germany}

\author{H. Guo}
\affiliation{Max-Planck-Institute for Chemical Physics of Solids, N\"{o}thnitzer Str. 40, D-01187 Dresden, Germany}

\author{Z. Hu}
\affiliation{Max-Planck-Institute for Chemical Physics of Solids, N\"{o}thnitzer Str. 40, D-01187 Dresden, Germany}

\author{T. S. Chan}
\affiliation{National Synchrotron Radiation Research Center (NSRRC), 101 Hsin-Ann Road, Hsinchu 30077, Taiwan}

\author{K. Nemkovski}
\affiliation{J\"{u}lich Centre for Neutron Science JCNS at Heinz Maier-Leibnitz Zentrum (MLZ), Forschungszentrum J\"{u}lich GmbH, Lichtenbergstra{\ss}e 1, 85748 Garching, Germany}

\author{A. C. Komarek}
\email[]{Alexander.Komarek@cpfs.mpg.de}
\affiliation{Max-Planck-Institute for Chemical Physics of Solids, N\"{o}thnitzer Str. 40, D-01187 Dresden, Germany}



\date{\today}

\begin{abstract}
Centimeter-sized single crystals of \slro\ were grown by the floating zone method at oxygen pressures of 20~bar. The quality of our single crystals was confirmed by X-ray Laue, powder and single crystal X-ray diffraction, neutron and X-ray absorbtion spectroscopy measurements. At $\sim$50\% electron doping we observe RhO$_3$ octahedral rotations of $\sim$8.2$^{\circ}$ within the octahedral basal plane which are incompatible with space group \emph{I4/mmm}. Our single crystal was further characterized by susceptibility, electrical transport and, finally, specific heat measurements showing a temperature dependent Debye temperature.
\end{abstract}


\maketitle


\section{Introduction}
Over the last decades strongly correlated transition metal oxides have attracted enormous attention \cite{rmp.66.445,zpb.64.189,cupratesbook,nature.358.136,cpb.22.087502,jpcb.104.5877}. In $4d$ transition metal oxides especially the discovery of the possible occurrence of spin-triplet $p$-wave superconductivity in ruthenates triggered a lot of research activity \cite{rmp.75.657,jcg.427.94}.
However, the rhodium oxides that also crystallize in the K$_2$NiF$_4$-type structure were less investigated so far. Besides of a comparative study of Sr$_2$RhO$_4$ single crystals isotypic with Sr$_2$RuO$_4$ mainly polycrystalline \slrox\ samples were studied over a wider doping range $x=0\sim1.0$ \cite{njp.8.175,jpsj.79.114719,prb.49.5591}. Regarding the recent observations of charge order and nano phase separation in the 3d cobaltate system La$_{2-x}$Sr$_{x}$CoO$_{4}$ \cite{nc.4.2449,nc.5.5731,rrl.9.580,JSNM} also the availability of single crystals of the 4d Rhodium analogue with hole-doping levels away from the end member Sr$_2$RhO$_4$ and, especially, around 50\% electron doping (\slro), would be of interest for future studies of the charge correlations.

In these rhodates Sr$_{2-x}$La$_{x}$RhO$_{4}$ the crystal structure changes from orthorhombic for $x=0\sim0.1$ to tetragonal for $x=0.1\sim0.6$ and back to orthorhombic for $x=0.6\sim1.0$. A doping dependent metal-insulator transition and a complex magnetic behavior that may be connected to the structural changes have been observed \cite{prb.49.5591}. Furthermore, a quasi-two-dimensional Fermi liquid state has been reported in floating zone grown single crystals of the undoped parent compound \sro\ \cite{njp.8.175,jpsj.79.114719}. More recently, unconventional magnetism has been reported in Ga or Ca doped \lsro\ \cite{prb.90.144402}. In these systems the occupation of the intermediate spin (IS) state ($S=1$) of the Rh$^{3+}$ ions might be the origin of this unconventional magnetic behavior. Note, that the low spin state (LS) ($S=0$) is commonly expected for the Rh$^{3+}$ ions.

In the present work, we report the growth of cm-sized high quality single crystals of 50\%\ electron doped \sro, i.e. \slro, by the flux feeding floating zone (FFFZ) method. When comparing with the expected value for a 50$\%$ Rh$^{3+}$ ($S=0$) and 50$\%$ Rh$^{4+}$ ($S=1/2$) solid solution, we find an enhancement of the effective magnetic moment (mainly) in the in-plane direction. Moreover, a temperature dependent Debye temperature can be observed for \slro. We discuss a possible thermal occupation of higher spin states of the $Rh$ ions as a possible mechanism that might be responsible for the interesting physics revealed in this system.

\section{Results and Discussion}

The \slro\ single crystal was grown following the procedure for the growth of Sr$_{2-x}$Ba$_{x}$RuO$_4$ \cite{jcg.427.94}. First, powders of SrCO$_3$ (99.99$\%$, Alfa Aesar), La$_2$O$_3$ (99.99$\%$, Alfa Aesar), and Rh$_2$O$_3\cdot x$H$_2$O (Rh-$\%=81.08\%$, Alfa Aesar) were mixed together and ground thoroughly before reaction at 1200\,$^oC$ in air for 24\,hours. The resulting powder was packed into latex tubes and pressed under $\sim100$\ MPa hydrostatic pressure into rods of $\sim6$\,mm diameter and $\sim140$\,mm ($\sim$35\,mm) length for the feeding (seed) rod. The obtained feed and seed rods were sintered at 1300\,$^oC$ for 48\,hours. Similar to the growth of ruthenates \cite{jcg.427.94} an excess of 5\% Rh$_2$O$_3$ needs to be added into the feed rod (with nominal composition Sr$_{1.5}$La$_{0.5}$Rh$_{1.1}$O$_x$) in order to serve as self-flux and in order to compensate the heavy evaporation during the growth. These rods were, then, mounted in a high pressure optical mirror furnace (High Pressure Crystal Growth Furnace, Scientific Instrument Dresden GmBH) equipped with one single Xenon lamp. A stable growth was maintained with a mixture of Ar and O$_2$ gas with a ratio of 1:1 flowing with a speed of 1\,L/min at a pressure of 20\,bar. The growth speed was set to 7.5\,mm/h. The feed and seed rod were counter rotated with a speed of 30~rotations per minute. The growth direction of this single crystal is roughly the [1~1~0]-direction. After floating zone growth the several cm-sized as-grown crystal with 6-7~mm in diameter was annealed at 800\,$^oC$ in flowing O$_2$ for 7\,days in order to remove any possible oxygen vacancies \cite{jpsj.79.114719}.

For powder XRD measurements parts of the grown single crystals have been ground into fine powders. The XRD measurements have been performed with a $2\theta$ step of $0.01^{o}$ with Cu $K_{\alpha1}$ radiation on a \emph{Bruker D8 Discover A25} powder X-ray diffractometer.
The \emph{FullProf} program package \cite{fullprof} was used for Le Bail fits.

Neutron scattering experiments were performed at the DNS spectrometer \cite{DNS} at the Heinz Maier-Leibnitz Zentrum
using an incident neutron wavelength of $\lambda$~$=$~4.2\AA.

Using Mo $K_{\alpha}$ radiation single crystal X-ray diffraction measurements have been performed on a \emph{Bruker D8 VENTURE} single crystal X-ray diffractometer  equipped with a bent graphite monochromator for about 3$\times$ intensity enhancement and a \emph{Photon} CMOS large area detector.
A spherical sample with roughly 100~$\mu$m diameter has been measured and a multi-scan absorption correction has been applied to the data (minimum and maximum transmission:  0.6203 and  0.7510 respectively). For space group \emph{I4/mmm} 42677 reflections (h: -10~$\rightarrow$~10, k: -7~$\rightarrow$~10 and l: -34~$\rightarrow$~31) have been collected with an internal R-value of 3.62\%, a redundancy of 67.53 and 99.21\% coverage up to sin($\Theta$)/$\lambda$~=~1.359).
For the final structure refinement with space group \emph{P2$_1$/c} the integration was redone and the refinement was based on a set of 80499 reflections (h: -18~$\rightarrow$~18, k: -14~$\rightarrow$~13 and l: -31~$\rightarrow$~34) have been collected with an internal R-value of 3.15\%, a redundancy of 14.48, 100\%  coverage up to 2$\Theta$~=~97.34$^{\circ}$ and 99.1\% coverage up to 2$\Theta$~=~144.94$^{\circ}$). Within this refinement four different twin domains with the following volume fractions and twin laws have been refined:  V$_1$~=~0.171(8),  V$_2$~=~0.316(4),  V$_3$~=~0.091(4),  V$_4$~=~0.421(4), M$_2$~=~(0.000  2.000  0.000, 0.500  0.000  0.000,  0.500  1.000 -1.000), M$_2$~=~(1.000  0.000  0.000, 0.000 -1.000  0.000,  1.000  0.000 -1.000) and M$_3$~=~(0.000 -2.000  0.000, 0.500  0.000  0.000, -0.500 -1.000  1.000).
The obtained structural parameters are listed in table~\ref{SXRDTB}.
The\emph{ Jana2006 }program suite \cite{Jana} was used for all crystal structure refinements.

The valence state of Rh was identified by X-ray absorption spectroscopy (XAS) using bulk sensitive fluorescence yield at the Rh-L$_3$ edge measured at the 16A beamline at the National Synchrotron Radiation Research Center in Taiwan.

Using a SQUID magnetometer (MPMS, Quantum Design) the magnetization was measured as a function of temperature with magnetic field applied along both crystallographic [110] and [001] directions. For resistivity measurements, the crystal was cut into a bar shape with typical dimensions of $\sim1\times1\times2$\,mm$^3$. Four probe contacts were made by connecting Au wires using silver paste and the measurements were performed by using a Physical Property Measurement System (PPMS, Quantum Design). The crystal was cut into a typical dimension of $\sim2\times2\times0.5$\,mm$^3$ along the $a$, $b$ and $c$ axes and the Specific heat was measured by using the relaxation method in our PPMS system.
Note, that all directions for susceptibility and resistivity measurements are given in tetragonal notation.

\subsection{Structure}

\begin{figure}[!h]
\includegraphics[width=0.85\columnwidth]{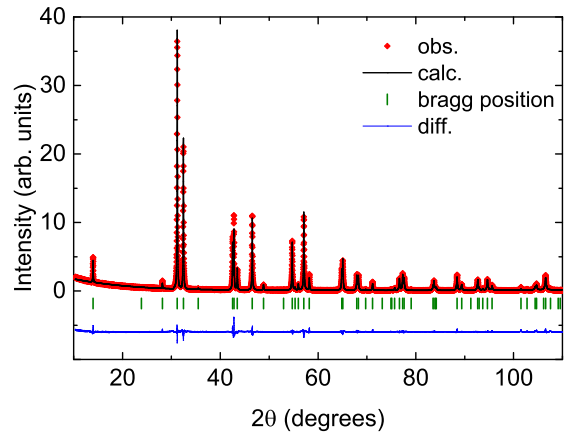}
\caption{(Color online) Powder X-ray diffraction pattern of the grown crystal \slro\ collected at room temperature. The solid line represents a Le Bail fit of the data using the FullProf software package \cite{fullprof}. The calculated Bragg peak positions according to space group I4/mmm are indicated by vertical bars and the difference between the experimental and calculated intensities are shown as solid blue line at the bottom.}
\label{XRD}
\end{figure}

\begin{figure}[!h]
\includegraphics[width=\columnwidth]{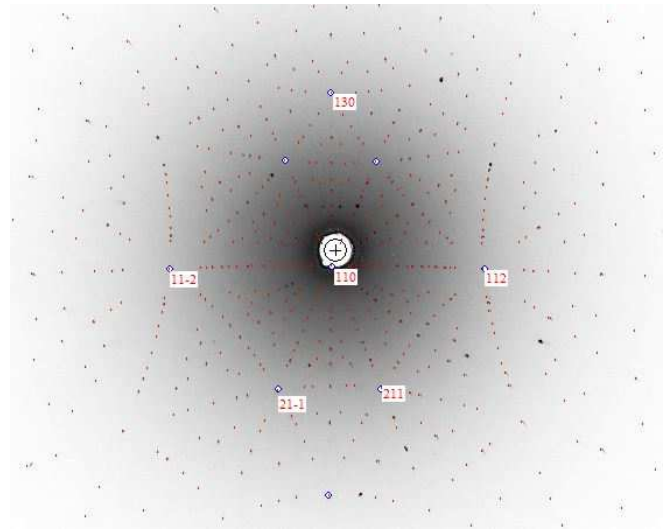}
\caption{(Color online) Laue diffraction pattern of \slro. The red spot is a simulation (space group \emph{I4/mmm}) using the software OrientExpress.}
\label{Laue}
\end{figure}

\begin{figure}[!h]
\includegraphics[width=\columnwidth]{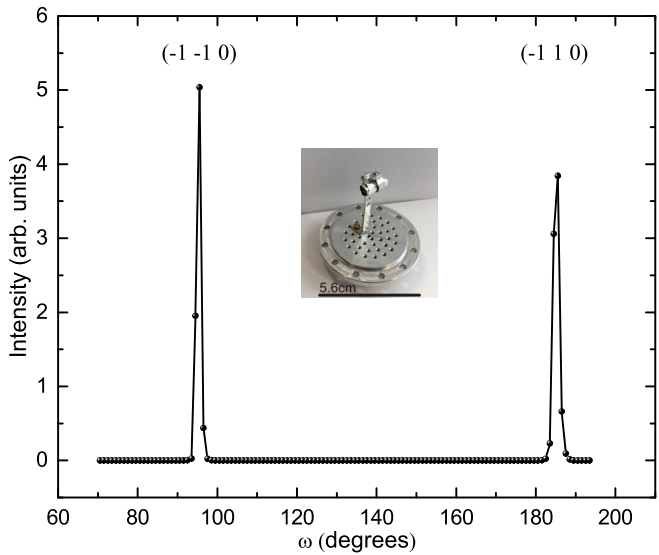}
\caption{(Color online) Large rocking scans measured in neutron scattering experiments confirming the single crystallinity of \slro. The inset shows a picture of the measured single crystal mounted in an Al sample holder and the black bar indicates the length scale (56~mm).}
\label{N}
\end{figure}

In Fig.~\ref{XRD} the powder X-ray diffraction (XRD) pattern of our \slro\ single crystal (crushed to powder) is shown. All peaks can be indexed with the tetragonal structure model with space group \emph{I4/mmm}. This is consistent with literature reports on polycrystalline powder samples \cite{prb.49.5591}. The lattice constants amount to $a=b=3.8989(8)$\,{\AA} and $c=12.667(3)$\,{\AA}. Laue diffraction measurements confirm the single crystallinity of our \slro\ crystals, see Figs.~\ref{Laue}. Moreover, neutron diffraction measurements that are penetrating and probing the entire single crystal further confirm the single crystallinity, see Fig.~\ref{N}.

\begin{figure}[!h]
\includegraphics[width=\columnwidth]{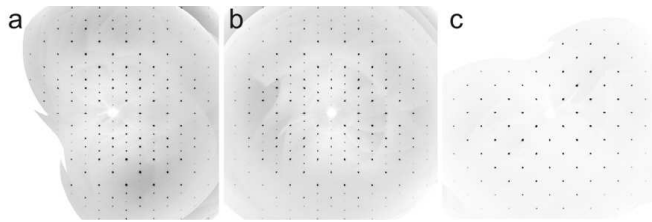}
\caption{(Color online) Single crystal X-ray diffraction intensities of \slro\ within the (a) $0KL$ (b) $H0L$ and (c) $HK0$ planes of reciprocal space (space group \emph{I4/mmm}).}
\label{scx}
\end{figure}

\par
\begin{figure}[!h]
\includegraphics[width=\columnwidth]{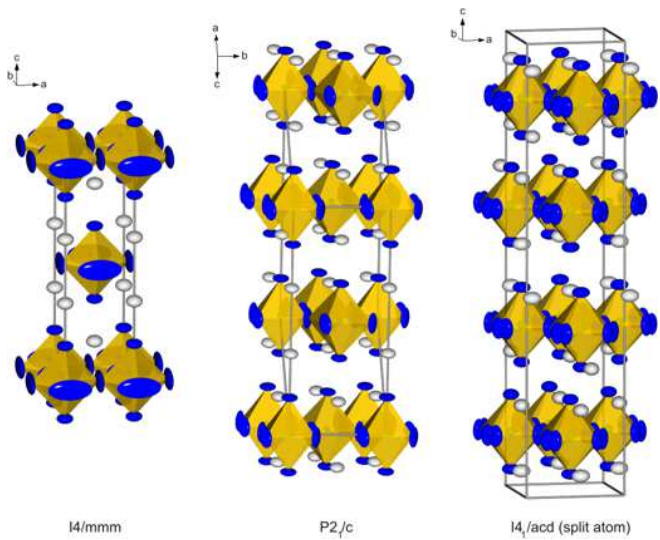}
\caption{(Color online) Representation of the crystal structure of \slro\ as obtained from refinements with space group \emph{I4/mmm},  \emph{P2$_1$/c} ($a$~=~6.9155\AA, $b$~=~5.5194\AA, $c$~=~12.682\AA, $\beta$~=~156.48$^{\circ}$ ) and with a split atom model based on space group \emph{I4$_1$/acd}. Grey/green/blue ellipsoids denote the probability ellipsoids of Sr(La)/Rh/O atoms. As can be seen, for the previously reported space group \emph{I4/mmm} of \slro\ the basal oxygen probability ellipsoids are extremely enlarged in direction perpendicular to the Rh-O bonds. This problem could be overcome only for the monoclinic symmetry of space group \emph{P2$_1$/c} or with a split atom model as reported in literature \cite{clock} for the end member Sr$_2$RhO$4$ with space group \emph{I4$_1$/acd}.}
\label{SXRD}
\end{figure}

 However, within a single crystal X-ray diffraction measurement - see figure~\ref{scx} - we find evidence for additional oxygen octahedral rotations around the $c$-axis.
Although less reflections are considered for the higher symmetric space group \emph{I4/mmm} a refinement with such a structure model yields enhanced values for the goodness of fit (GoF) and strongly elongated probability ellipsoids of the basal oxygen ions in a transversal direction to the Rh-O bonds,
see figure~\ref{SXRD}(a).
Therefore, we tried to overcome these problems by applying several known structure models for layered perovskites based on space groups \emph{Fmmm}, \emph{Bmeb}, \emph{P4$_2$/ncm}, \emph{I4$_1$/acd} and \emph{I4$_1$/a}. But the strongly elongated probability ellipsoids that indicate RhO$_3$ octahedral rotations around the $c$-axis remain essentially unchanged for all these space groups (including the subgroups \emph{I4$_1$} and \emph{C2/c}). Moreover, GoF and R-values stay always enlarged for all these structure models, see table \ref{SXRDTA}.
Especially, also the crystal structure of the end member Sr$_2$RhO$_4$ which crystallizes with the symmetry of space group \emph{I4$_1$/acd} \cite{EndMember} does not yield any satisfactory description of our single crystal X-ray data.

\begin{table}
   \begin{ruledtabular}
  \begin{tabular}{c|ccc|c}
   & GoF & R-value (\%) & R$_w$-value (\%) & largest U$_{ii}$ (\AA$^{2}$) \\
   \hline
   \emph{I4/mmm} & \textbf{9.85} & 2.90 & 6.70 & \textbf{0.133(6)} \\
   \emph{Fmmm}   & \textbf{7.43} & 2.91 & 6.69 & \textbf{0.076(14)} \\
   \emph{Bmeb}   & \textbf{5.43} & 3.57 & \textbf{13.54} &  \textbf{0.076(7)} $^{\dag}$  \\
   \emph{P4$_2$/ncm} & \textbf{5.18} & 4.64 & \textbf{13.73} & \textbf{0.065(6)} \\
   \emph{I4$_1$/acd} & 4.68 & 3.73 & \textbf{12.92} & \textbf{0.053(3)} \\
   \emph{I4$_1$/a} & 3.90 & 3.22 & \textbf{11.11} & \textbf{0.058(14)} \\
   \emph{I4$_1$} & 3.29 & 3.21 & \textbf{9.87} & \textbf{0.070(15)} \\
   \emph{Ibca} & 4.01 & 3.39 & \textbf{11.53} & \textbf{0.069(4)} \\
   \emph{C2/c} & 3.48 & 3.26 & \textbf{10.43} & \textbf{0.097(14)} \\
   \hline
   \emph{P2$_1$/c} & 1.97 & 2.62 & 6.68 & 0.018(1) \\
   \hline
   split atom: & & & & \\
   \emph{I4$_1$/acd}$^{\ddag}$  & 3.70 & 3.08 & \textbf{10.22} & 0.032(1) \\
   \hline
  \end{tabular} \end{ruledtabular}
  \caption{Refinement results of single crystal X-ray diffraction measurements of \slro. Here, the GoF and R-values obtained for refinements with different space groups are listed together with the largest value of any U$_{ii}$; the symbol $^{\dag}$ indicates that U$_{iso}$ was listed because only a refinement with isotropic displacement parameters for the oxygen atoms turned out to be stable (all heavier atoms were still refined with anisotropic ADP). The bold values are strongly enlarged values which are not in favor for the corresponding structure model. For all these space groups - apart from \emph{P2$_1$/c} -
   the basal oxygen ions have one extremely enlarged anisotropic displacement parameter U$_{ii}$.
   Finally, we were able to describe the crystal structure of \slro\ with space group \emph{P2$_1$/c}, see also figure~\ref{SXRD}.
   However, also a split atom model $^{\ddag}$ is able to overcome the highly enlarged U$_{ii}$. Nevertheless, the weighted R-value stays significantly larger than for the monoclinic solution \emph{P2$_1$/c}.
  }\label{SXRDTA}

\end{table}

Therefore, we started to search for solutions based on lower symmetric space groups. Finally, we were able to describe the crystal structure of \slro\ with  space group \emph{P2$_1$/c} and to obtain acceptable GoF and R-values, see table~\ref{SXRDTA}.
Especially, the extremely strong elongation of the basal oxygen ellipsoids in a direction transversal to the Rh-O bonds which was present in all other refinements (with anisotropic ADP) has almost vanished for the structure model based on the monoclinic space group \emph{P2$_1$/c}, see figure~\ref{SXRD}.
The \slro\ crystal structure contains RhO$_3$ octahedra that are elongated in apical direction (i.e. the Rh-O distances amount to $\sim$2.080~\AA\ in apical direction and $\sim$1.967(1)~\AA\ and $\sim$1.976(1)~\AA\ in basal direction).
Moreover, the basal Rh-O-Rh bond angles amount to $\sim$163.6(1)$^{\circ}$ and, thus, indicate basal octahedral rotations of about 8.2$^{\circ}$ whereas the apical oxygens are only tilted by less than 2$^{\circ}$.
Since mainly oxygen displacements are involved, these structural distortions are difficult to detect with powder X-ray diffraction which might explain why these distortions could not be observed before (figure~\ref{XRD} and Ref.~\cite{prb.49.5591}).
Moreover, this single crystal X-ray diffraction measurement is able to corroborate a close to stoichiometric composition of our single crystal, i.e. the refinement of La, Sr and Rh occupancies indicates the following composition: La$_{0.481(8)}$Sr$_{1.519(8)}$Rh$_{0.991(2)}$O$_4$.
The details of the crystal structure are listed in table~\ref{SXRDTB}.

\begin{table}
  \begin{ruledtabular}
  \begin{tabular}{l|cccccc}
    atom & occup. & x & y & z & & \\
   \hline
      Sr1 & 0.760(4) & 0.00237(17) & -0.00066(9) & 0.35698(9) \\
   La1 & 0.240(4)    & 0.00237(17) & -0.00066(9) & 0.35698(9) \\
   Rh1 & 0.9909(18) &  0 &  0 &  0  \\
   O1 & 1 & -0.5733(5) & 0.2855(2)  & -0.7879(5) \\
   O2 & 1 &  0.0052(16) & 0.0016(9)  & 0.1662(8) \\
   \hline
   \hline
   & & & & & & \\
    atom & U$_{11}$ (\AA$^{2}$) & U$_{22}$ (\AA$^{2}$) & U$_{33}$ (\AA$^{2}$) &  & &  \\
   \hline
   Sr1 &  0.00907(5) &  0.00949(6) &  0.00737(9)   &  & &  \\
   La1 &  0.00907(5) &  0.00949(6) &  0.00737(9)   &  & &  \\
   Rh1 &  0.00261(4) &  0.00297(4) &  0.00143(7)   &  & &  \\
   O1  &  0.0090(5)  &  0.0079(4)  &  0.0093(10)   &  & &  \\
   O2  &  0.0176(6)  &  0.0179(6)  &  0.0137(8)    &  & &  \\
   \hline
     atom &   U$_{12}$ (\AA$^{2}$) & U$_{13}$ (\AA$^{2}$) & U$_{23}$ (\AA$^{2}$) & U$_{iso}$ (\AA$^{2}$)  &   & \\
   \hline
   Sr1 &  -0.00033(6)&  0.00781(6) & -0.00026(10) &   0.0076(3)    & \\
   La1 &  -0.00033(6)&  0.00781(6) & -0.00026(10) &   0.0076(3)    & \\
   Rh1 &  0.00001(7) &  0.00179(4) &  0.00002(10) &   0.0026(2)    & \\
   O1 &   -0.0060(2) &  0.0067(6)  & -0.0048(6)   &   0.015(3)     & \\
   O2 &   0.0000(6)  &  0.0152(6)  & -0.0001(9)   &   0.013(3)     & \\
   \hline
   \hline
  \end{tabular} \end{ruledtabular}
  \caption{Refinement results of single crystal X-ray diffraction measurements of \slro. Here, the structural parameters of the refinement with space group \emph{P2$_1$/c} ($a$~=~6.9155\AA, $b$~=~5.5194\AA, $c$~=~12.682\AA, $\beta$~=~156.48$^{\circ}$ ) are listed, see also figure~\ref{SXRD}(b).
  Goodness of fit, R- and weighted R-values amount to 1.97, 2.62\% and 6.68\% respectively.}\label{SXRDTB}

\end{table}

Finally, we would like to note that there was a structural study on Sr$_2$RhO$_4$ which proposes the occurrence of microdomains in Sr$_2$RhO$_4$ with clockwise and anticlockwise rotations of the RhO$_3$ octahedra around the $c$-axis that are allowed in space group \emph{I4$_1$/acd} \cite{clock}. A structure model with a statistical (random) rotation of the RhO$_3$ octahedra around the $c$-axis was able to describe the neutron diffraction data of Sr$_2$RhO$_4$ properly \cite{clock}. This structure model has been realized by splitting the basal oxygen site according to the two different possible rotation directions - clockwise and anticlockwise - of the RhO$_3$ octahedra which are allowed within space group \emph{I4$_1$/acd} \cite{clock} (but not e.g. in space group \emph{I4/mmm} that has been previously reported for \slro\ \cite{prb.49.5591}). The occupation of these split oxygen sites O2 and O2' indicates the total sizes of domains with clockwise and anticlockwise rotation.
In principle, this split atom model with the absence of fully long range ordered rotations of the RhO$_3$ octahedra could also be a good description of the crystal structure of \slro\ because the major shortcoming of any crystal structure model that is listed in table \ref{SXRDTA} is the extreme elongation of the probability ellipsoids of the basal oxygen ions which strongly indicates a wrong description of the RhO$_3$ octahedral rotations around the $c$-axis for all these models.
Hence, we also refined a split atom model similar as in Ref.~\cite{clock} for \slro\ - see table~\ref{SXRDTA}.
Besides our monoclinic solution (\emph{P2$_1$/c}) also this split atom model with statistical rotations of the RhO$_3$ octahedra around the $c$-axis \cite{clock} does not suffer from these extremely enlarged probability ellipsoids of the basal oxygen ions, see also Fig.~\ref{scx}. In this split atom model the RhO$_3$ octahedra are statistically rotated by $\pm$$\sim$8.0$^{\circ}$ around the $c$-axis which is close to the value of the octahedral rotations in our monoclinic solution ($\sim$8.2$^{\circ}$). However, the weighted R-values are significantly better for our monoclinic structure model, see table~\ref{SXRDTA}. Hence, the monoclinic solution appears to be more likely, and, maybe also the parent compound Sr$_2$RhO$_4$ needs to be re-analyzed.
One main qualitative difference of these two structure models is the absence of any RhO$_3$ octahedral tilts (i.e. rotations either of the apical or of the basal oxygen ions around any rotational axis perpendicular to the tetragonal $c$-axis) for the split atom model based on space group \emph{I4$_1$/acd}. Such tilts are known to occur in layered perovskites with K$_2$NiF$_4$ structure (e.g. for space group \emph{Bmeb}) and these kind of tilts are allowed for the monoclinic space group \emph{P2$_1$/c} in addition to the rotations of the RhO$_3$ octahedra around the tetragonal $c$-axis.
All these octahedral rotations are important for an understanding of the electronic structure of \slro, similar as was pointed out for the end member Sr$_2$RhO$_4$ in Ref.~\cite{elec}.

\subsection{X-ray absorption spectroscopy }
\begin{figure}[!h]
\includegraphics[width=1\columnwidth]{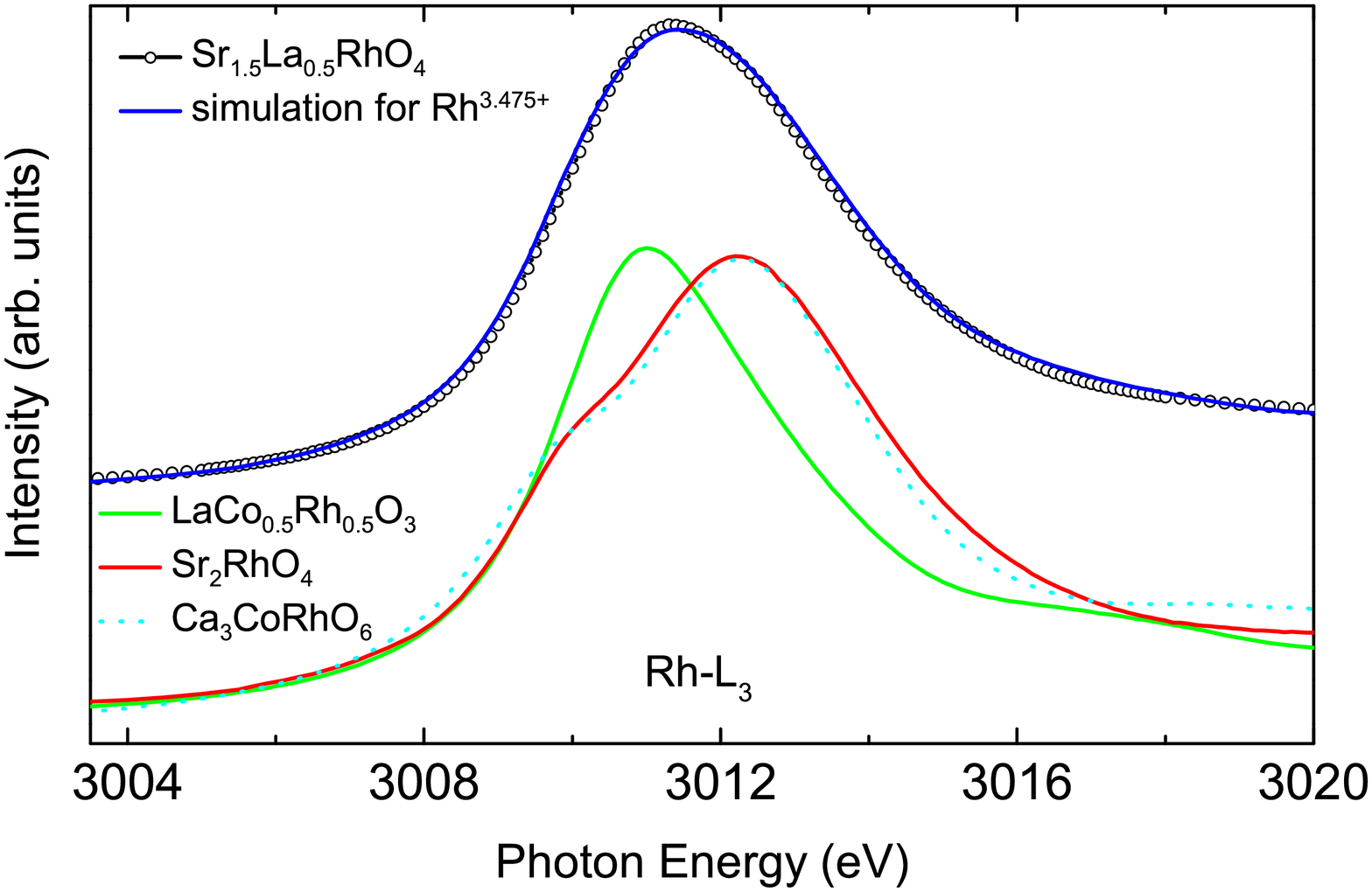}
\caption{(Color online) X-ray absorption spectroscopy measurements. Below the spectrum of \slro\ (black circles) the Rh$^{3+}$ and Rh$^{4+}$ references LaCo$_{0.5}$Rh$_{0.5}$O$_3$ (green line)) and Sr$_2$RhO$_4$ (red line) are shown. Also the spectrum of Ca$_3$CoRhO$_6$ (dashed cyan line) as another  Rh$^{4+}$ reference is shown. A superposition of the Rh$^{3+}$ and Rh$^{4+}$ reference spectra (blue line) with weights of 52.5\% and 47.5\% respectively is able to model the XAS spectrum of \slro\ (black circles) convincingly. Hence the Rh valence in our \slro\ single crystal amounts to $\sim$3.475+ which is close to the nominal value of 3.5+ in this compound. Note, that the deviation from its nominal value is of the order of the estimated error bar of $\sim$2\% in these kind of measurements.}
\label{XAS}
\end{figure}

It is well known that XAS spectra at the 4d transition metal L$_{2,3}$ edge are very sensitive to the valence state - an increase of the valence state of the metal ion by one causes a shift of the XAS L$_{2,3}$ spectra by one or more eV towards higher energies \cite{XAS}.
The XAS spectrum of \slro\ is just in between that of the Rh$^{3+}$ and Rh$^{4+}$ references LaCo$_{0.5}$Rh$_{0.5}$O$_3$ and Sr$_2$RhO$_4$, respectively.
A superposition of 52.5\%\ of the Rh$^{3+}$ and 47.5\%\ of the Rh$^{4+}$ reference spectra is nicely able to reproduce the XAS spectrum of \slro,
thus, indicating a Rh valence state of roughly 3.475+ which is close to its nominal value of 3.5+ and which would correspond to a small oxygen deficiency of the order of $\delta$~$\sim$~-0.0125 in our Sr$_{1.5}$La$_{0.5}$RhO$_{4+\delta}$ single crystal - see Fig.~\ref{XAS}. However, the deviation of the measured valence from its nominal value is also of the order of the estimated error bar of $\sim$2\% in our XAS measurements.

\subsection{Magnetic properties}

\begin{figure}[!t]
\includegraphics[width=\columnwidth]{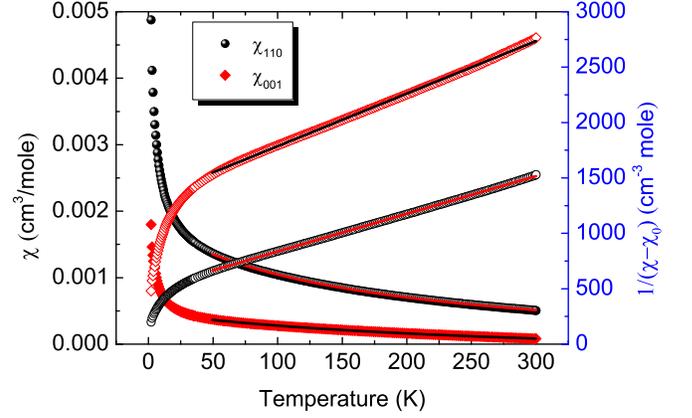}
\caption{(Color online) Temperature dependence of the magnetic susceptibilities $\chi(T)$ (filled symbols) of \slro\ measured with an external field of \textbf{H}\,=\,5000\,Oe applied along the crystallographic [110] and [001] directions. Inverse magnetic susceptibilities $1/(\chi-\chi_0)$ (open symbols) are shown to the right axis. Solid lines between 50\,K to 300\,K are Curie-Weiss-like fits to the experimental data.}
\label{MT}
\end{figure}

The magnetic susceptibility $\chi(T)$ measured in an external field of \textbf{H}\,=\,5000\,Oe is shown in Fig.~\ref{MT}. $\chi_{[110]}$ and $\chi_{[001]}$ are the magnetic susceptibilities measured with \textbf{H}\,$\parallel$\,[110]-direction and \textbf{H}\,$\parallel$\,[001]-direction, respectively. We observe an moderate anisotropy of $\chi_{[110]}/\chi_{[001]}=2\sim6$ over the whole measured temperature range. No hysteresis between zero-field cooled (ZFC) and field cooled (FC) curves can be observed. The solid line in Fig.~\ref{MT} is a fit of $\chi(T)$ containing a Curie-Weiss and a temperature independent term $\chi_0$:
\begin{equation}
\label{curieweiss}
\chi(T) = \chi_0 + \frac{C}{T-\theta}
\end{equation}
with the Curie constant $C=N\mu^2_{eff}/(3k_B)$, Weiss temperature $\theta$, number of spins $N$, Boltzmann constant $k_B$ and the effective magnetic moment per formula unit $\mu_{eff}$. The resulting fitting parameters for $\chi_{[110]}(T)$ amount to $C_{[110]}=0.295(3)$\,cm$^3$K/mole, $\chi_{0,[110]}=-1.48(5)\times10^{-4}$\,cm$^3$/mole, and $\theta_{[110]}=-147(1)$\,K. For $\chi_{[001]}(T)$ these values amount to $C_{[001]}=0.211(5)$\,cm$^3$K/mole, $\chi_{0,[001]}=-2.81(6)\times10^{-4}$\,cm$^3$/mole, and $\theta_{[001]}=-277(5)$\,K. The effective magnetic moments calculated from these parameters are $\mu_{eff,[110]}=1.54(1)$\,$\mu_B$ and $\mu_{eff,[001]}=1.30(2)$\,$\mu_B$ for [110]-direction and [001]-direction, respectively. The determined effective magnetic moments here are slightly smaller than that reported earlier from polycrystalline samples, $\mu_{eff}=1.738$\,$\mu_B$ \cite{prb.49.5591}.
In \slro\ only the Rh ions with oxidation state 3.5+ contribute to the magnetism. Due to the localized nature of \slro\ (see our resistivity measurements below), it is reasonable to assume a statistical distribution of Rh$^{4+}$ ($S=1/2$) and Rh$^{3+}$ ($S=0$) ions, thus, yielding $\langle S(S+1)\rangle=3/8$. Assuming a g-factor of 2, the effective magnetic moment amounts to  $\mu_{eff,calc.}=1.225$\,$\mu_B$. This calculated value is somewhat smaller than the above two experimentally obtained values from our Curie-Weiss fit. This discrepancy is either indicative for a thermal population of an intermediate or high spin (HS) state of the Rh ions \cite{prb.90.144402} or for a somewhat larger g-factor \cite{prb.90.144402,jltp.159.11}.

\subsection{Resistivity}
\begin{figure}[!t]
\includegraphics[width=0.8\columnwidth]{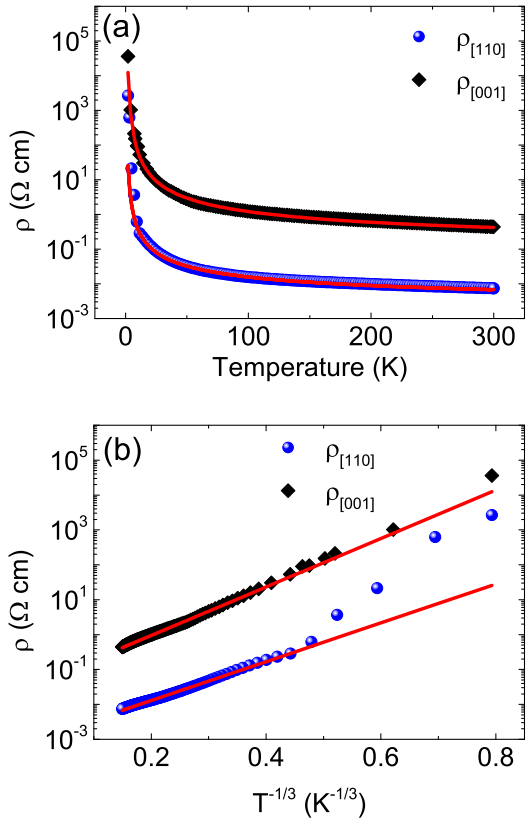}
\caption{(Color online) (a) Resistivity as a function of temperature and (b) resistivity vs $T^{-1/3}$ on a logarithmic scale for \slro\ measured with the electrical current flow along the crystallographic [110]/[001]-directions as indicated in the legends. Solid lines represent fits to experimental data points with the two dimensional Mott variable-range hopping theory in the temperature range between 50\,K to 300\,K (see text).}
\label{RT}
\end{figure}

The in-plane and out-of-plane resistivity $\rho_{[110]}(T)$ and $\rho_{[001]}(T)$ of \slro\ was measured with an applied electrical current along the crystallographic [110] and [001]-directions respectively, see Fig.~\ref{RT}~(a). The room temperature values of the electrical resistivity amount to $\rho_{[110]}(300\,K)=0.75\times10^{-2}\,\Omega$\,cm and $\rho_{[001]}(300\,K)=0.44\,\Omega$\,cm. With decreasing temperature, the resistivity measured in both directions increases to values of five orders of magnitude larger. This reveals the insulating nature of \slro. We fitted the experimental data between 50\,K to 300\,K with the two dimensional Mott variable-range hopping model \cite{ap.21.785}, $\rho(T)=\rho_0 exp[(T_0/T)^{(1/(d+1))}]$, where $d=2$ for a two dimensional layered system, see Fig.~\ref{RT}~(a) (solid line). This is different to studies on polycrystalline samples in literature where plural types of conduction mechanisms were observed, i.e. neither a simple activated mechanism nor a variable-range hopping mechanism was reported for \slro\ \cite{prb.49.5591}. The resulting parameters from our fits to $\rho_{[110]}(T)$ and $\rho_{[001]}(T)$ amount to $\rho_{0,[110]}=9.9(2)\times10^{-4}$\,$\Omega$\,cm, $T_{0,[110]}=2098(37)$\,K and $\rho_{0,[001]}=3.88(9)\times10^{-2}$\,$\Omega$\,cm, $T_{0,[001]}=4078(58)$\,K, respectively.
For clarity, we plot the resistivity data as a function of $T^{-1/3}$ in Fig.~\ref{RT}~(b). Solid lines were calculated using the fitting parameters above.
Only at low temperatures, $\sim T\leq15$\,K, the data points start to deviate from the theoretical model, thus, suggesting that electron-electron interactions and/or magnetic scattering start to play some role at low temperatures.

\begin{figure}[!t]
\includegraphics[width=\columnwidth]{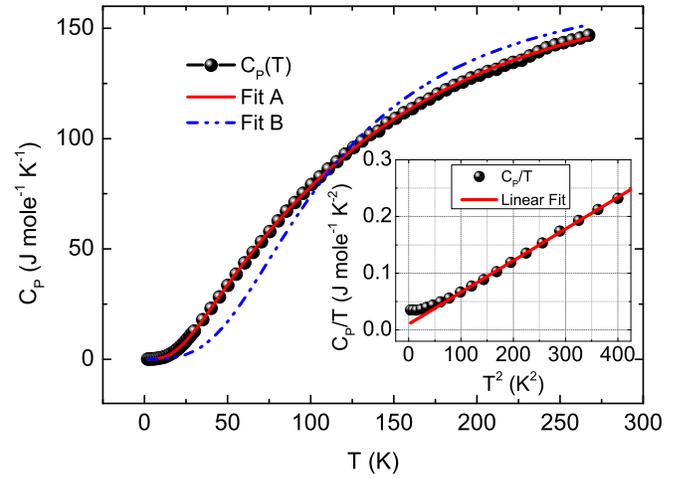}
\caption{(Color online) Specific heat $C_P$ of \slro\ single crystal as a function of temperature measured in zero magnetic field. The red solid and blue dashdotted curves are fits to the experimental data using the Debye lattice Specific heat model with temperature dependent and constant Debye temperatures, respectively (see text). Inset shows $C_P/T$ versus $T^2$ in the low temperature range, $T\leq20$\,K. The straight red line represents a linear fit to the data between 10\,K to 20\,K.}
\label{CP}
\end{figure}

\subsection{Specific heat}
In Fig.~\ref{CP} the zero field specific heat $C_P(T)$ of our \slro\ single crystals is shown. No obvious anomaly can be observed in the entire temperature range which is in agreement with our magnetization and resistivity measurements.
Our measured values of $C_P$ at high temperatures, e.g. $C_P(267$\,K$)=147$\,J\,mole$^{-1}$\,K$^{-1}$, are significantly lower than the classical Dulong-Petit prediction of the lattice specific heat $C_V=3nR=175$\,J\,mole$^{-1}$\,K$^{-1}$, where $n$ is the number of atoms per formula unit and $R$ is the molar gas constant. This suggests that \slro\ has a Debye temperature $\Theta_D$ distinctly higher than 267\,K, which is consistent with published data for the undoped parent compound Sr$_2$RhO$_4$ ($\Theta_D$(Sr$_2$RhO$_4$)$\,=421$\,K) \cite{jpsj.79.114719}. As shown in the inset of Fig.~\ref{CP}, $C_P(T)/T$ versus $T^2$ exhibits a linear region at low temperatures (between 10\,K to 20\,K). Therefore, we fitted the data in this range by the formula $C_P(T)/T=A+\beta T^2$, where the weak upturn below $\sim10$\,K and the small but non-vanishing first term $A=1.0(1)\times10^{-2}$\,J\,mole$^{-1}$\,K$^{-1}$ could arise from the slowing down of critical spin fluctuations. The second term corresponds to the low temperature limit of the Debye lattice Specific heat with fitted value of $\beta=5.59(5)\times10^{-4}$\,J\,mole$^{-1}$\,K$^{-4}$. From this value of $\beta$ we can estimate the Debye temperature using the expression $\Theta_D=(12\pi^4nR/5\beta)^{1/3}=289.8(9)$\,K. This value is much smaller than one would expect from the high temperature limit data, suggesting that the system temperature has a non-negligible effect on the Debye temperature of \slro.
To quantify this effect, we tried to fit the entire data set using the Debye lattice Specific heat model \cite{Kittelbook}
\begin{equation}
\label{DebyeCv}
C_V(T) = 9R(\frac{T}{\Theta_D})^3\int_0^{\Theta_D/T}\frac{x^4e^x}{(e^x-1)^2}dx.
\end{equation}
In the fit, we used the analytic Pad\'{e} approximant function given in Ref.~\cite{prb.85.054517} that accurately represents the above integral. As shown by the blue dashdotted line ($\Theta_D=456(6)$\,K) in Fig.~\ref{CP}, a reasonable fit can not be obtained with a constant Debye temperature. In order to estimate the temperature dependence of the $\Theta_D(T)$, we assume that the system exhibits a ground state with $\Theta_{D0}$ at base temperature. With increasing temperature, the system is thermally activated to a state separated by $\Delta E$ with probability of $e^{-\frac{\Delta E}{k_BT}}$ and we further assume that this activated state has a simple form of Debye temperature $\Theta_{D1}=\Theta_{D0}+Constant$. Hence, it follows that
\begin{equation}
\label{ThetaDT}
\Theta_D(T)=\Theta_{D0}(1+Ae^{-\frac{\Delta E}{k_BT}}).
\end{equation}
As shown by the red solid curve in Fig.~\ref{CP}, an excellent fit can be obtained by our simple two-states model with parameters of $\Theta_{D0}=291(4)$\,K, $\Delta E=4.9(2)$\,meV and $A=1.14(2)$.
The origin of this behaviour is subject of future studies.

\section{Conclusions}

In summary, we have grown cm-sized single crystals of \slro\ by the FFFZ method. We have characterized our crystals by X-ray, neutron, susceptibility, electrical transport and specific heat measurements. We find indications either for an enhanced in-plane effective magnetic moment or for an anisotropic g-factor and we observe a temperature dependent Debye temperature for our \slro\ single crystal. Moreover, we have strong indications that the crystal structure of \slro\ can not be described with space group \emph{I4/mmm} as reported in literature. After comparing different models the best description of the crystal structure can be achieved with space group \emph{P2$_1$/c}. The symmetry lowering arises from both, RhO$_3$ octahedral rotations of $\sim$8.2$^{\circ}$ around the former tetragonal $c$-axis and octahedral tilts of slightly less than 2$^{\circ}$.
\par
\begin{acknowledgments}
We thank L. H. Tjeng for helpful discussions.
\end{acknowledgments}

\end{document}